\documentclass[final]{raa06}           
\usepackage{graphicx,times}             
\usepackage{natbib}
\usepackage{amssymb,amsmath}
\bibpunct{(}{)}{;}{a}{}{,}
\usepackage[colorlinks=true, citecolor=blue]{hyperref}%

\usepackage{orcidlink}

\usepackage{graphicx,kantlipsum,setspace}
\usepackage{caption}
\usepackage{graphics,epsf}
\usepackage{amsmath}                
\usepackage{amsfonts}               
\usepackage{amssymb}                
\usepackage{epsfig}                 
\usepackage{appendix}
\usepackage{graphicx}
\usepackage{float}
\usepackage{color}
\usepackage{multirow}
\usepackage{colortbl}
\usepackage[para,online,flushleft]{threeparttable}
\usepackage{xcolor}

\hypersetup{citecolor=blue, 
            linkcolor=red, 
            menucolor=blue, 
            urlcolor=blue}  

\newcommand{\km}{{~\rm km}}
\newcommand{\s}{{~\rm s}}

\newcommand{\erg}{{~\rm erg}}

\begin{document}

   \title{Supernova 1987A’s keyhole: A long-lived jet-pair in the final explosion phase of core-collapse supernovae
}

   \volnopage{Vol.0 (20xx) No.0, 000--000}      
   \setcounter{page}{1}          

   \author{Noam Soker, \orcidlink{0000-0003-0375-8987}
    }

   \institute{Department of Physics, Technion, Haifa, 3200003, Israel;   {\it   soker@technion.ac.il}\\
\vs\no
   {\small Received~~20xx month day; accepted~~20xx~~month day}}

\abstract{I further study the manner by which a pair of opposite jets shape the `keyhole' morphological structure of the core-collapse supernova (CCSN) SN 1997A, now the CCSN remnant (CCSNR) 1987A. By doing so, I strengthen the claim that the jittering-jet explosion mechanism (JJEM) accounts for most, likely all, CCSNe. The `keyhole' structure comprises a northern low-intensity zone closed with a bright rim on its front and an elongated low-intensity nozzle in the south. This rim-nozzle asymmetry is observed in some cooling flow clusters and planetary nebulae that are observed to be shaped by jets. I build a toy model that uses the planar jittering jets pattern, where consecutive pairs of jets tend to jitter in a common plane, implying that the accreted gas onto the newly born neutron star at the late explosion phase flows perpendicular to that plane. This allows for a long-lived jet-launching episode. This long-lasting jet-launching episode launches more mass into the jets that can inflate larger pairs of ears or bubbles, forming the main jets' axis of the CCSNR that is not necessarily related to a possible pre-collapse core rotation. I discuss the relation of the main jets' axis to the neutron star's natal kick velocity. 
\keywords{stars: massive -- supernovae: general -- stars: jets -- ISM: supernova remnants}  }

 \authorrunning{N. Soker}            
\titlerunning{The point-symmetric structure of SN 1987A}  
   
      \maketitle

\section{Introduction} 
\label{sec:intro}

The formation of a neutron star (NS) by the collapsing inner core of a massive star releases $\simeq {\rm few} \times 10^{53} \erg$ of gravitational energy; neutrino-anti-neutrino pairs carry most of this energy (e.g., \citealt{Janka2012}). There is no consensus on the processes that utilize a small fraction of this gravitational energy to explode the rest of the star and power a core-collapse supernova (CCSN) with typical explosion energies of $E_{\rm exp} \approx 10^{49} -10^{52} \erg$ (e.g., \citealt{Burrows2013}). Recent studies discuss either the delayed-neutrino explosion mechanism (e.g., \citealt{Andresenetal2024, BoccioliFragione2024, BoccioliRoberti2024, Burrowsetal2024b, Burrowsetal2024, GhodlaEldridge2024, JankaKresse2024, Maunderetal2024, MatsumotoTakiwakiKotake2024, Schneideretal2024, WangTBurrows2024}, limiting the list to 2024), or the jittering-jets explosion mechanism (JJEM; e.g., 
\citealt{BearSoker2024, ShishkinSoker2024, Soker2024NA1987A, Soker2024N63A, Soker2024Rev, Soker2024CFs, Soker2024PNSN1987A, WangShishkinSoker2024}).

In the second half of 2023 (see \citealt{Soker2024Rev} for a review), there was a breakthrough as far as the JJEM is concerned by identifying point-symmetric morphologies in several CCSN remnants (CCSNRs), as is the expectation of the JJEM (e.g., \citealt{BearSoker2023RNAAS}). The identification of several symmetry axes that form a \textit{point-symmetric wind-rose} in the iconic CCSNR Cassiopeia A \citep{BearSoker2024} and the identification of a rim-nozzle asymmetry (Section \ref{sec:Motivation}) in the ejecta of SN 1987A \citep{Soker2024CFs, Soker2024PNSN1987A} substantially strengthen the claim that most, and probably all, CCSN are exploded by jets and in the frame of the JJEM (for the discussion of these eight CCSNRs see \citep{Soker2024CFs}.
In the JJEM, in most (but not all) cases, the jets are not expected to be relativistic (e.g., \citealt{PapishSoker2011, PapishSoker2014Planar, GilkisSoker2014}; Section \ref{sec:Motivation}). In a later study, \cite{Piranetal2019} mention the possible importance of non-jittering relativistic jets in some CCSNe, mainly striped-envelope CCSNe. \cite{Izzoetal2019} find outflow velocity of $\simeq 10^5 \km\s^{-1}$ in SN 2017iuk, which is associated with gamma-ray burst GRB 171205A, and conclude that the jets also play a role in energizing the CCSN associated with the GRB.

The morphologies of many CCSNRs contain two opposite ears (e.g., \citealt{GrichenerSoker2017ears, Soker2023SNRclass}). Some point-symmetric CCSNRs also have a main jet-axis in addition to other axes, SNR 0540-69.3  \citep{Soker2022SNR0540}, Vela SNR \cite{Soker2023SNRclass}, and SN 1987A \citep{Soker2024CFs}. The main ejecta structure inside the equatorial ring (which is a circumstellar material) of SN 1987A is an elongated structure with a low-intensity hole in the north and an elongated low-intensity zone in the south (Section \ref{sec:Motivation}). This structure of SN 1987A is sometimes referred to as the `keyhole'. The axis of the `keyhole' is not along the symmetry axis of the equatorial ring, which most likely was formed by a binary interaction. Namely, the axis of the main jet-axis (the `keyhole') is not along the angular momentum of the binary progenitor system. The question is, what determines the main jet axis of CCSNRe when it occurs? This study's topic is to answer this question in the JJEM frame. I start by reviewing some of the relevant properties of the JJEM (Section \ref{sec:TypicalValues}). In Section \ref{sec:Motivation}, I discuss the `keyhole' of SN 1987A (or already a CCSNR: SNR 1987A) as the motivation for this study.
In Section \ref{sec:ToyModel} I build a toy model to explain the formation of a main jet-axis in the frame of the JJEM. I summarize in Section \ref{sec:Summary}.   

\section{Typical values for typical CCSNe} 
\label{sec:TypicalValues}

With the JJEM's great success last year in accounting for point-symmetric CCSN remnants, the challenge now is to determine the properties of the jittering jets during the explosion process and simulate the star's explosion. This will be achieved in full only with highly demanding sophisticated magnetohydrodynamical simulations. Currently, there are only estimates of the relevant properties; some are more robust, and some are crude. 

In Table \ref{Tab:Table1}, I list the values of some of the parameters of jittering jets and list the arguments for these parameters and the relevant references.  The energy of the jets that inflate the ears of CCSNRs and the number of jet-launching episodes that it implies were derived from analyzing observations of CCSNRs \citep{Bearetal2017,  GrichenerSoker2017ears}. The launching velocity of the jets is taken to be the escape velocity from the NS as is observed in other astrophysical systems that launch jets (e.g., \citealt{Livio2004}; supported by \citealt{Guettaetal2020} who argue that most CCSNe have no signatures of relativistic jets). The fraction of the accreted mass the jets carry is also taken as in other astrophysical systems.    
The bottom two rows, in red, list the two new properties of the present study that I present in Section \ref{sec:ToyModel}. 
The values in the table and what follows are for FeCCSNe, namely, CCSNe, where an iron-rich core collapses. For electron-capture CCSNe, the jets are less energetic, and the jet-driven explosion phase might be much longer, i.e., minutes to even a few hours rather than seconds \citep{WangShishkinSoker2024}.  
\begin{table*}
\begin{center}
  \caption{Typical parameters of jittering jets in FeCCSNe}
    \begin{tabular}{| p{5.0cm} | p{3.8cm}| p{7.0cm}| }
\hline  
\textbf{Property} & \textbf{Values} & \textbf{Justification}  \\
\hline  
Jet launching velocity & $v_j \simeq 10^5 \km \s^{-1}$ & The escape velocity from the NS as in most other astrophysical types of objects, e.g., [Li04]$^\&$ \\ 
\hline  
Relative energy of an ear pair to total explosion energy $E_{\rm exp}$&  $\epsilon_{\rm ears} \simeq 0.03-0.2 $ & Studies of ears in CCSNRs [GrSo17] [BGS17] \\ 
\hline  
Energy in one pair of jets$^{\#}$ &  $E_{\rm 2j} \simeq 10^{49} - 2 \times 10^{50} \erg $  & Studies of ears in CCSNe and their $E_{\rm exp}$ [BGS17] \\ 
\hline  
Total number jittering-jets pairs  & $N_{\rm 2j} \simeq 5-30$  &  $N_{\rm 2j} \simeq (\epsilon_{\rm ears})^{-1}$ [PaSo14a]  \\
\hline  
Mass in one pair of jets & $m_{\rm 2j} \simeq 10^{-4}-0.002 M_\odot$  &  $m_{\rm 2j} \simeq 2E_{\rm 2j}/v^2_j $  [PaSo14a] \\
\hline  
Accreted mass from a disk/belt in one episode & $m_{\rm 1acc} \simeq 0.001-0.02 M_\odot$  &  $m_{\rm 2j} \approx 0.1 m_{\rm 1acc}$ as with other astrophysical types of objects (e.g., young stellar objects, e.g., [Ni18])  \\
\hline  
Duration of main explosion phase$^{\#}$ & $\tau_{\rm exp} \simeq 0.5-10 \s$ & 1-several free-fall times from exploding layer at $\approx 3000 \km$ [PaSo14b], [ShSo21] \\
\hline  
Duration of one jet-launching episode & $\tau_{\rm 2j} \approx 0.01-0.3 \s$ & $ \la \tau_{\rm exp} / N_{\rm 2j}$  \\
\hline  
Accretion disk's relaxation time & 
$\tau_{\rm vis} \approx 0.01-0.1 \s$ & 10-100 times orbital period on the surface of the NS [PaSo11], [So24]  \\
\hline  
{\textcolor{red}{\textbf{Toy-model early (main) phase parcel-accretion-rate}}} & $\dot N_{\rm p,E} \approx 10-50 \s^{-1}$  &  Number of convective cells accreted divided by explosion time [GiSo15]; equation (\ref{eq:dotNpe}) \\
\hline  
{\textcolor{red}{\textbf{Toy-model late parcel-accretion-rate}}} & $\dot N_{\rm p,L} \approx 1-5 \s^{-1}$  &  Here (Section \ref{sec:ToyModel}) \\
\hline  
     \end{tabular}
  \label{Tab:Table1}\\
\end{center}
\begin{flushleft}
\small 
Notes: The last two rows are quantities I introduce in this study and elaborate on in Section \ref{sec:ToyModel}. 
\newline
$^{\#}$ The values in the table are for CCSNe of collapsing iron-rich core (FeCCSNe). Jets of electron capture supernovae are less energetic, and the explosion might last for minutes to a few hours \citep{WangShishkinSoker2024}. 
\newline 
$^\&$ \cite{Izzoetal2019} reported the possible indication for jets at $\simeq 10^5 \km \s^{-1}$ in SN 2017iuk associated with GRB 171205A.  
\newline
References: [BGS17]: \cite{Bearetal2017}; [GiSo15]: \cite{GilkisSoker2015};[GrSo17]: \cite{GrichenerSoker2017ears}; [Li04]: \cite{Livio2004}; [Ni18]: \cite{Nisinietal2018}; [PaSo11]: \cite{PapishSoker2011}; [PaSo14a]: \cite{PapishSoker2014D2}; [PaSo14b]: \cite{PapishSoker2014Planar}; [ShSo14]: \cite{ShishkinSoker2021}; [So24]: \cite{Soker2024N63A}. 
\end{flushleft}
\end{table*}

In the JJEM, the newly born NS accretes gas with stochastically varying angular momentum in magnitude and direction (e.g., \citealt{GilkisSoker2014}). If there is a pre-collapse core rotation, the stochastic variation is around this axis (e.g., \citealt{Soker2023gap}). Namely, it is not completely stochastic at a full solid angle. In the toy model that I built in this study (Section \ref{sec:ToyModel}), the accreted gas is decomposed into parcels of gas. One or more gas parcels combine to form an accretion disk (or accretion belt) of one jet-launching episode. Namely, the total number of parcels in the toy model is larger than that of the jet-launching episodes. 

The new parameters I introduce here are the accretion rate of gas parcels during two phases that I define here: the early (main) and late phase. 
The early explosion phase is the phase when the NS launches most of the jets. It is the main explosion phase. 
I estimate the early parcel accretion rate as follows. The typical mixing length in the exploding layer, i.e., the core layer that, when accretes, explodes the star, is ${\rm ML} \simeq 0.3r-0.4r$, where $r\simeq 2000- 3000 \km$ is the radius of the exploding layer  (e.g., \citealt{ShishkinSoker2021}). In a layer of width ML there are $N{\rm conv} \approx 4\pi r^2 / \pi ML^2 \approx 20-40$ convective cells (e.g., \citealt{GilkisSoker2015}). I crudely estimate the early parcel accretion rate to be this number divided by the typical explosion time in the early (main) phase. As this is the main phase, its duration is not much shorter than the typical explosion time $\tau_{\rm exp}$ which is about the free-fall time from the exploding layer to the center (see Table \ref{Tab:Table1}). I therefore crudely estimate 
\begin{equation}
\dot N_{\rm p, E} \approx  \frac{N{\rm conv}}{\tau_{\rm exp}} \approx 10-50  \s^{-1} .
\label{eq:dotNpe}
\end{equation}
It is important to note that this is a crude estimate and that several gas parcels combine during the early phase to form one jet-launching episode. The estimated number of jet-launching episodes is $N_{\rm 2j} \simeq 5-30$ (see Table \ref{Tab:Table1}), from which I estimate that $N{\rm conv}/N_{\rm 2j} \approx 2-10$ parcels combine to give one jet-launching episode in the early (main) phase of the explosion process. In Section \ref{sec:ToyModel}, I speculate on the parcel accretion rate at the late phase, when the accretion flow is not completely random, and the accretion rate decreases. 

\section{Motivation} 
\label{sec:Motivation}

The viscous timescale of the accretion disk is not much shorter or might be longer than the lifetime of the intermittent accretion disk of one jet-launching episode (\citealt{PapishSoker2011}; Table \ref{Tab:Table1}). This ratio is  $1 \lesssim (\tau_{\rm 2j} / \tau_{\rm vis}) \lesssim 10$  \citep{Soker2024N63A}.  Because of the short lifetime of the intermittent accretion disk, the intermittent accretion disk is not expected to settle into a thin accretion disk completely \citep{PapishSoker2011}, and the two sides of the disk might not be exactly equal to each other, implying that the two opposite jets that the disk launches are not exactly equal in their properties \citep{Soker2024N63A}. In particular, one jet in a pair can be more energetic than the counter jet. One possible morphological outcome is that the stronger jet breaks out from the bubble (or lobe) it inflates while the weaker jet does not. The stronger jet opens a nozzle out of the bubble/lobe, which appears as a lower-emission zone through the front of the bubble/lobe. On the other hand,  the weaker jet compresses a dense cap on the front of the bubble it inflates, which appears as a high-intensity rim (arc). This \textit{rim-nozzle asymmetry} is observed in the X-ray images of a few cooling flow clusters and some planetary nebulae and the CCSNRs G107.7-5.1 and SN 1987A \citep{Soker2024CFs, Soker2024PNSN1987A}.  In cooling flows and some planetary nebulae, such pairs of bubbles are observed to be inflated by jets. The JJEM attributes such structures in CCSNRs, including ears, to jets that exploded the star.
Note that asymmetry in the ambient gas density can also allow one jet to break out from the bubble (on the lower-density side) while arresting the other jets on the other side (of higher ambient density), even when the jets have equal properties. 

\begin{figure}
\begin{center}
\includegraphics[trim=0.1cm 15.1cm 0.0cm 0.0cm ,clip, scale=0.53]{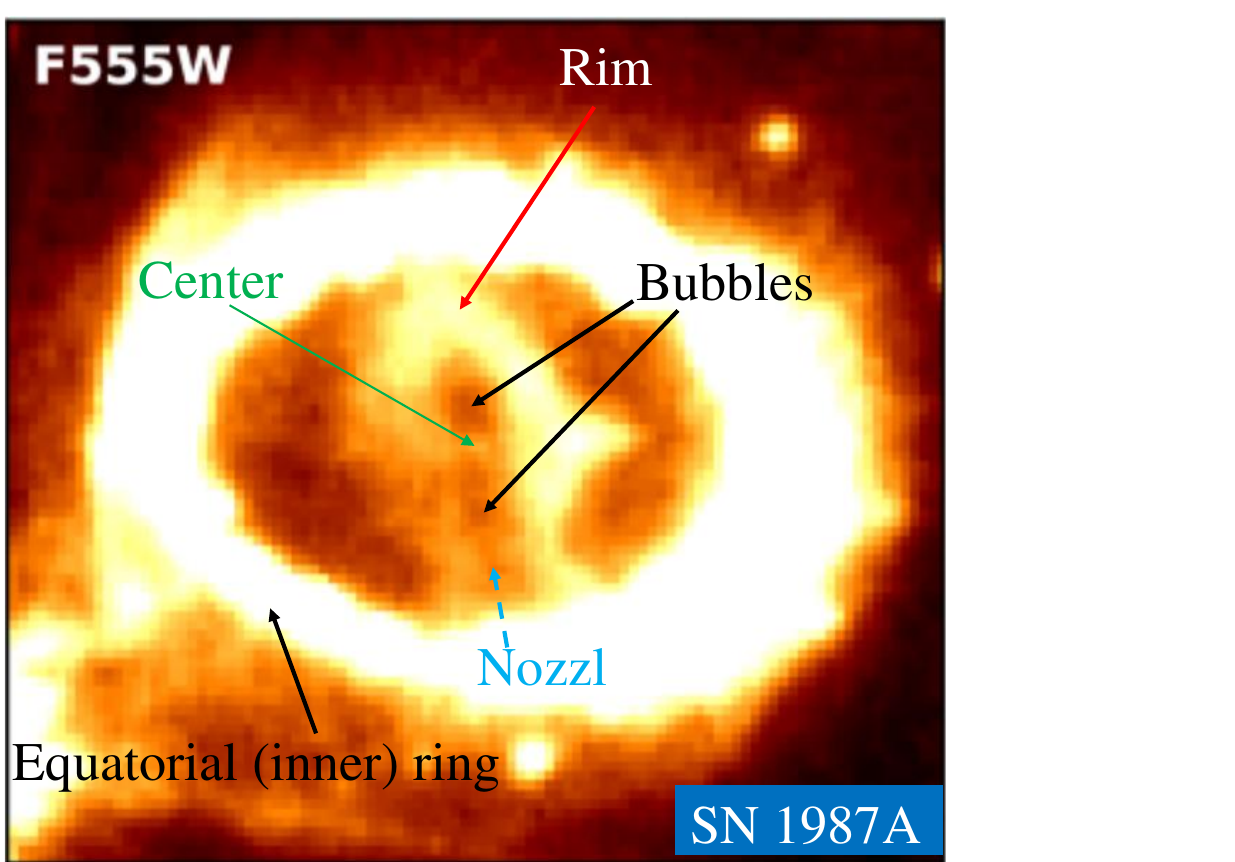}
\end{center}
\caption{An HST/WFC3 image of SN 1987A in one filter from \cite{Rosuetal2024}; north is up and east to the left. The `keyhole' is the north-south elongated bright structure inner to the equatorial ring.  In \cite{Soker2024CFs}, I identified a northern bubble (upper thick-black arrow) with its front rim (red arrow) and a southern nozzle (dashed-light-blue arrow). 
The rim-nozzle asymmetry is similar to the structures of some cooling flow clusters and planetary nebulae that are observed to be shaped by jets. A jet that broke out from the southern bubble formed the nozzle. In the north, the jet did not break out due to a weaker jet or (less likely) a denser ambient gas.  
}
\label{Fig:SN1987A}
\end{figure}

In \cite{Soker2024CFs}, I attributed the bright elongated structure of the ejecta of SN 1987A, which is sometimes termed the `keyhole' (e.g., https://esawebb.org/images/SN1987a-2/), to such a jet-shaped rim-nozzle asymmetry. Observations over the years reveal this `keyhole' structure, in particular some recent high-quality JWST (e.g., \citealt{Arendtetal2023}) and HST (e.g., \citealt{Rosuetal2024}) observations. The recent observations by \cite{Rosuetal2024} caused me to claim a rim-nozzle asymmetry to explain the keyhole-like morphology. In Figure \ref{Fig:SN1987A}, I present one image from  \cite{Rosuetal2024}, to which I added marks related to the JJEM. 
The observations are from 12,980 days after the explosion. The `keyhole' is the north-south elongated structure of the ejecta that is mainly smaller than the equatorial ring (which is a pre-explosion circumstellar gas). Two low-intensity zones compose the `keyhole'. In the north, I identified \citep{Soker2024CFs} this zone as a bubble, based on comparison to bubbles (X-ray cavities) in cooling flow clusters. In the south, I identified the elongated low-intensity zone as a bubble with a nozzle radially outward, as I marked in Figure \ref{Fig:SN1987A}. 

In this study, I address the question of why the axis of the rim-nozzle line (the long axis of the `keyhole') is close to being in the plane of the equatorial ring. If this axis is the angular momentum of the pre-collapse core, then why is this axis not along the symmetry axis of the equatorial ring and the two outer rings of SN 1987A? A simple model might be that the binary companion that shaped the three rings (as thought to be the case in similar morphologies of planetary nebulae) also spun up the core. This seems not to be the case. Moreover, many CCSNRs have an elongated large-scale structure with two opposite ears along this axis, suggesting a shaping by a long-lived pair of two opposite jets according to the JJEM (e.g., \citealt{Bearetal2017, GrichenerSoker2017ears, Soker2023SNRclass}). I suggest an explanation for the launching of one to a few long-lived jets at the late phase of the explosion process, according to the JJEM. For that, I turn to build a toy model.

\section{The toy model} 
\label{sec:ToyModel}

I build the following phenomenological toy model in the frame of the JJEM. I consider the accretion of gas parcels with identical mass $m_{\rm p}$ and a specific angular momentum magnitude of $\vert \overrightarrow{j}_{\rm p}\vert $. Note that this is the angular momentum after amplifying the seed perturbations from the pre-collapse convective core layers by instabilities above the NS. The gas parcels form an accretion disk (or an accretion belt) around the NS. I take the $z$ axis along the disk's angular momentum axis. The specific angular momentum $z$-component of the parcel of gas is  
\begin{equation}
j_z=j_{\rm p} \cos \theta, . 
\label{eq:jzE}
\end{equation} 
where $\theta$ is the angle between $\overrightarrow{j}_{\rm p}$ and the $z$ axis.  
Consider the main JJEM phase, i.e., the early phase, in a case with no pre-explosion core rotation. In that case, the direction of $\overrightarrow{j}_{\rm p}$ is random, so the probability for an angle $ 0 \le \theta < \pi$ is $(1/2) \sin \theta d \theta$. 
The average of the absolute value of many parcels of gas is 
\begin{equation}
\vert \overline{j}_{z,{\rm E}}\vert =\frac{1}{2}\int_0^{\pi} \vert j_{\rm p} \cos \theta \vert \sin \theta d \theta =\frac{1}{2} j_{\rm p}  . 
\label{eq:Barjz}
\end{equation}

Let the accretion rate of gas parcels per time unit at the early phase when jittering is fully random be $\dot N_{\rm p,E}$. Again, each parcel has the same mass and value of $j_{\rm p}$. In the time $\tau_{\rm E}$, the number of accreted parcels is $\tau_{\rm E} \dot N_{\rm p, E}$, some have negative and some positive value of $\overline{j}_{z,{\rm E}}$. The typical specific angular momentum of the accreted gas in that period (take it to be positive) is 
\begin{equation}
j_{z,{\rm acc, E}} \approx \frac{1}{\sqrt{\tau_{\rm E} \dot N_{\rm p, E}}} \vert \overline{j}_{z,{\rm E}} \vert . 
\label{eq:jAccE}
\end{equation}
The condition to maintain an accretion disk is that this value be larger than the minimum value of a test particle orbiting at the surface of the NS $j_{\rm disk}$, i.e., $ j_{z,{\rm acc, E}} > j_{\rm disk}$. 
The typical lifetime of the accretion disk in the simple toy model, with the aid of equation (\ref{eq:Barjz}) and for $\tau_{\rm E} \dot N_{\rm p, E}$> 1, is 
\begin{equation}
\tau_{\rm E} \approx 
\left( \frac{j_{\rm p}}{2 j_{\rm disk}} \right)^2 
\frac{1}{\dot N_{\rm p, E}} . 
\label{eq:TauE}
\end{equation}

As \cite{PapishSoker2014Planar} showed, the jittering pattern might change after two or more jet-launching episodes to have the jets' axes sharing a plane. The argument goes as follows. A given pair of opposite jets accelerates material outward along and near the two jets' directions. Therefore, at later times, the material is more likely to be accreted from directions perpendicular to the jets' axis. This flow type holds for the next jet-launching episode with a jets' axis inclined to the first jet's axis. The two axes define a plane, so the inflow is perpendicular to that plane. The accretion inflow is practically along an axis perpendicular to the plane of jittering (from both sides of that plane). The angular momentum of the inflowing gas is perpendicular to its inflow velocity, therefore, in the plane of the two earlier jets' axis. Namely, the jets' axis, which is the angular momentum axis, of the next jet-pair will tend to be in the same plane as the previous jets' axes. This is termed a planar jittering-jets pattern.  Large angular momentum fluctuations of the accreted gas might break the planar jittering pattern. 

The planar jittering-jets pattern has two main effects on the accreted gas that launches the next pairs of jets. The first is the planar pattern. Namely, the jets' axes do not jitter in a full solid angle of $4 \pi$, but rather the jets' axes jitter in a plane. The probability for a jets' angle in the plane and $ 0 \le \theta < \pi$ is $(1/\pi) d \theta$. The average of the absolute value of many parcels of gas is 
\begin{equation}
\vert \overline{j}_{z, {\rm L}}\vert =\frac{1}{\pi}\int_0^{\pi} \vert j_{\rm p} \cos \theta \vert d \theta =\frac{2}{\pi} j_{\rm p}  . 
\label{eq:jzL}
\end{equation}
Repeating the steps leading to equation (\ref{eq:TauE}) gives the disk lifetime during the planar jittering-jets phase 
\begin{equation}
\tau_{\rm L} \approx 
\left( \frac{2 j_{\rm p}}{\pi j_{\rm disk}} \right)^2 
\frac{1}{\dot N_{\rm p, L}} , 
\label{eq:TauL}
\end{equation}
where $\dot N_{\rm p, L}$ is the parcel accretion rate in the late phase during the planar jittering-jets pattern. 

The second effect of the planar jittering-jets pattern is that, due to the much smaller inflow-solid angle, the accretion rate substantially decreases at these late times, $\dot N_{\rm p, L} \ll \dot N_{\rm p, E}$.
The ratio of accretion disk lifetime at a late time during a planar jittering-jets phase to that at early phases of the explosion process when the jittering is (almost) fully random in all directions is given by the ratio of equation (\ref{eq:TauL}) to equation (\ref{eq:TauE}). This reads   
\begin{equation}
\frac{\tau_{\rm L}}{\tau_{\rm E}} \approx 
\left( \frac{4}{\pi} \right)^2 
\frac{\dot N_{\rm p, E}}{\dot N_{\rm p, L}} . 
\label{eq:TauRatio}
\end{equation}
The first term on the right-hand side contributes a factor of 1.62. It will likely be smaller as the accretion at late times is not exactly from an axis but rather from a cone. The main effect of the planar jittering-jets pattern that might prolong the life of the accretion disk is in reducing the mass accretion rate onto the accretion disk around the newly born NS. 

Although in Table \ref{Tab:Table1} I listed values of several determined parameters of the JJEM, there are some parameters that the JJEM does not have yet. These include the opening angle of the jets, the value of $j_{\rm p}$ that the toy model uses, and the parcel accretion rate at the late explosion phase $N_{\rm p, L}$. I expect the value of $N_{\rm p, L}$ to vary in a large range of $N_{\rm p, L} \simeq 1 \s^{-1}$ to $N_{\rm p, L} \simeq  N_{\rm p, E}$ when there is no planar jittering-jets pattern. 
To have a long-lived pair of jets (or even two or three pairs) that shape a prominent structure in the CCSNR, the value of $N_{\rm p, L}$ should be small. This is the value I estimated in Table \ref{Tab:Table1}. Despite these unknowns, the toy model does have merit in explaining the formation of a major jet axis that is not necessarily determined by pre-collapse core rotation. When there is a rapid pre-collapse core rotation, it might determine the main axis of the descendant CCSNR.  

Overall, the toy model shows the following behavior of some CCSNRs when the progenitor core did not have a rapid pre-collapse rotation. In the main (early) jet-launching phase, the intermittent accretion disks (or belts) live for short times because new parcels of gas accreted at a high rate rapidly change the angular momentum direction and magnitude. In the late phase, the time difference from one parcel of gas to the next is typically much longer. An accretion disk can live longer and launch more mass along a fixed axis. This pair of jets will be more energetic than the average and, therefore, shape a prominent structure in the CCSNRs. In other words, in the late phase, a jet-launching episode terminates when the disk is depleted of its gas rather than being destroyed by the following parcels of gas in the early phase. I suggest this accounts for prominent elongated structures, like pairs of ears, in some CCSNRs.  

\section{Summary} 
\label{sec:Summary}

This short study is another step in exploring the properties and outcomes of the JJEM.  The specific goal was to account for the presence of one, or even two or three, main jet axes that shape the descendant CCSNR. The direct motivation (Section \ref{sec:Motivation}) comes from the recent identification of the `keyhole' structure of SN 1987A ejecta as inflated by a pair of jets \citep{Soker2024CFs} and from the direction of its long axis, i.e., the axis of the jets that shaped the `keyhole.' In the southern bubble of the `keyhole', the jets broke out to form a nozzle, while the northern jet compressed a rim. Figure \ref{Fig:SN1987A} presents this rim-nozzle asymmetry. The jets' axis, which is the long axis of the `keyhole,' is not along the symmetry axis of the equatorial ring. This suggests that the angular momentum of the progenitor binary system, which likely shaped the equatorial and two outer rings, did not determine the axis of the `keyhole'. I built a toy model (Section \ref{sec:ToyModel}) to explain the presence of one to a few energetic pairs of jets at the end of the explosion process, even when the pre-collapse core does not rotate.  

The toy model uses the planar jittering jets pattern, where consecutive pairs of jets tend to jitter in a common plane. I showed that this allows for one (or possibly two or three) long-liven jet-launching episodes at the late phase of the explosion process when the mass accretion rate by the newly born NS substantially decreases. Such a long-lasting jet-launching episode launches more mass into the jets that can inflate larger pairs of ears or bubbles. This is the main jets' axis of the CCSNR. I estimated the duration of this long-lasting jet-launching episode to be $\approx 0.2-1 \s$; this estimate requires further study. I do not claim that I explained all properties of the `keyhole'. This requires further study and exploration of the full three-dimensional structure of the `keyhole' and the rest of the ejecta. I limited myself to account for the direction of its long axis in the frame of the JJEM. 

I list the two parameters the toy model introduces, which are the accretion rate of gas parcels in the early (main) and late phases of the explosion process, at the bottom of Table \ref{Tab:Table1}. The other rows in the table give the typical parameters of the JJEM that were derived in earlier studies. 

The formation of a main jets' axis in CCSNRs is also related to the natal kick velocity of the NS. Studies \citep{BearSoker2018, Soker2022SNR0540, BearSoker2023RNAAS} examined the angle between the natal kick velocity of the NS and the main jets' axis (its projection on the plane of the sky) and deduced that the NS natal kick velocity avoids small angles relative to the main jets' axis. 
The explanation within the JJEM frame for this avoidance of small kick-jet angle (e.g., \citealt{Soker2023kick}) uses the gravitational tug-boat mechanism. In the tug-boat mechanism (e.g., \citealt{Wongwathanaratetal2013}), one to several ejecta clumps gravitationally pull and accelerate the NS to its natal kick velocity. According to the JJEM, the final pair of jets, which mostly shape the outer ejecta, prevent the formation of dense clumps along their propagation direction, hence no acceleration of the NS along this jets' axis. Another possibility is that the gas parcels that feed the final accretion disk and the group of clumps that gravitationally accelerate the NS have a common source. This implies that the acceleration direction of the NS is in the plane of the last accretion disk, and because the jets are perpendicular to the accretion disk, the natal kick velocity is at a large angle to the axis of the last jets. In both of these explanations, a more energetic pair of jets is compatible with more massive ejected clumps that gravitationally accelerate the NS. 

This study adds to recent arguments based on CCSNR morphological features favoring the JJEM as the main explosion mechanism of most, likely all, CCSNe. Neutrino heating does take place, but only in boosting the jet-driven explosion \citep{Soker2022boosting}; jittering jets play the main role in exploding CCSNe.  
 
\section*{Acknowledgements}
I thank an anonymous referee for helpful comments. A grant from the Pazy Foundation supported this research.

\end{document}